# Optoelectronic manifestation of orbital angular momentum driven by chiral hopping in helical Se chains

Bumseop Kim[1,2,†], Dongbin Shin[3,4,†], Seon Namgung[2], Noejung Park[1,2], Kyoung-Whan Kim[5,*], Jeongwoo Kim[6,*]

[1]Graduate School of Semiconductor Materials and Devices Engineering, Ulsan National Institute of Science and Technology, Ulsan 44919, Korea

[2]Department of Physics, Ulsan National Institute of Science and Technology, Ulsan 44919, Korea

[3]Department of Physics and Photon Science, Gwangju Institute of Science and Technology, Gwangju 61005, Korea

[4]Max Planck Institute for the Structure and Dynamics of Matter and Center for Free Electron Laser Science, Hamburg 22761, Germany

[5]Center for Spintronics, Korea Institute of Science and Technology, Seoul 02792, Korea

[6]Department of Physics, Incheon National University, Incheon 22012, Korea

[†]These authors contributed equally: Bumseop Kim, Dongbin Shin

*corresponding author e-mail
kwk@kist.re.kr; kjwlou@inu.ac.kr

## Abstract

Chiral materials have garnered significant attention in the field of condensed matter physics. Nevertheless, the magnetic moment induced by the chiral spatial motion of electrons in helical materials, such as elemental Te and Se, remains inadequately understood. In this work, we investigate the development of quantum angular momentum enforced by chirality using static and time-dependent density functional theory calculations for an elemental Se chain. Our findings reveal the emergence of an unconventional orbital texture driven by the chiral geometry, giving rise to a non-vanishing current-induced orbital moment. By incorporating spin-orbit coupling, we demonstrate that a current-induced spin accumulation arises in the chiral chain, which fundamentally differs from the conventional Edelstein effect. Furthermore, we demonstrate the optoelectronic detection of the orbital angular momentum in the chiral Se chain, providing an alternative to the interband Berry curvature, which is ill-defined in low dimensions.

## TOC

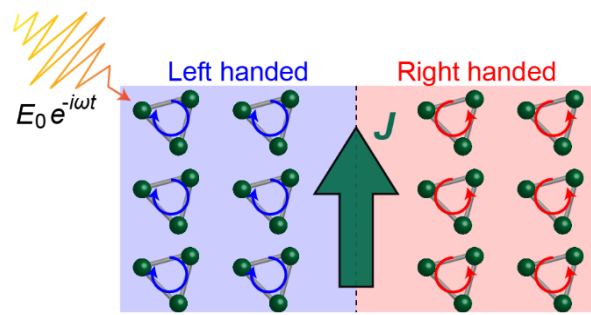

# Introduction

The concept of chirality has emerged as an essential factor in condensed matter physics, playing a crucial role in determining nontrivial band topology and exotic spin texture.[1-5] The intimate relation between momentum and spin, which is enforced by chirality, has been studied in a variety of systems, including the chiral anomaly of Weyl semimetals,[6-9] high-fold chiral fermions,[2, 10-13] chiral domain walls,[14] and magnetic skyrmions.[15] Moreover, the chiral spatial motion dictated by helical geometry is closely related to the dynamics of spin angular momentum (SAM) as demonstrated by the chirality-induced spin selectivity and the chiral magnetic effect.[3, 5, 16-18] Intriguingly, the observed spin magnetic moment induced by the chiral motion in helical materials is significantly greater, by one to three orders of magnitude, than that predicted by the classical solenoid model.[19] This large discrepancy indicates the existence of a strong quantum-mechanical interplay between the motion of electron and its SAM in chiral systems.

Elemental Te and Se represent a promising platform for exploring the intricacies of the interwoven orbital motion and magnetism due to its helical atomic structure and strong spin-orbit coupling (SOC).[20, 21] The trigonal-stacked crystal structure of Te and Se consists of one-dimensional chains with three-fold screw rotation symmetry,[22, 23] thereby enabling facile control over size and thickness in low dimensions.[24-26] In Te nanowires, a large chirality-dependent magnetoresistance is observed along the helical chain, and its induced magnetism is several orders of magnitude larger than in non-chiral systems.[27-29] The sizable magnetic moment is expected to be the result of the Edelstein effect, which refers to the spin accumulation induced by a flowing charge current.[30, 31] As evidenced by the Rashba-Edelstein effect, spin-momentum locking is required for the current-induced spin accumulation.[32] In the case of Te chains, the radial spin texture in momentum space (see Figure 1b for example), a unique spin-momentum locking of chiral chains, is a key ingredient responsible for the large current-induced magnetism.[21, 33, 34]

Nevertheless, as implied by the discrepancy between the spin magnetic moment predicted by the classical solenoid theory and the experimental result in chiral materials, the fundamental physics underlying the distinctive spin-momentum locking and the microscopic mechanism governing the large current-induced magnetism in the chiral chain systems have not been sufficiently investigated. Although previous helical model studies presented a close correlation between the radial spin texture and the inherent geometry of helical chains,[19, 23, 35-37] the quantum mechanical origin underpinning this association remains elusive. Given that orbital angular momentum (OAM) can also be created by an external electric field in helical chains, it is imperative to clarify its contribution to the induced magnetic moment and its interaction with SAM through strong SOC.

In this study, we examined the spontaneous development of coupled OAM and SAM induced by the helical geometry in chiral Se chains. We also investigated the quantum mechanical behavior of excited carriers in non-equilibrium states under the influence of diverse external fields using real-time time-dependent density functional theory (TDDFT). It is found that the chiral hopping between the *p*-orbitals of an individual Se chain produces a notable orbital polarization in the vicinity of the Fermi level. This chiral OAM gives rise to a distinct spin polarization via strong SOC, which corresponds to the radial spin texture observed in bulk. Our simulation demonstrates that the current-induced magnetism in chiral chains is primarily driven by the quantum superposition of the orbital states, which is clearly distinct from the conventional Edelstein effect that arises from the slight shift of the Fermi surface. We also present a potentially useful method for detecting the intriguing OAM embedded in the chiral Se chain using the one-to-one correspondence between OAM and photocurrent. We unveil how the spin and orbital degrees of freedom are entangled in a helical chain, the quantum mechanical counterpart of a classical solenoid, and provide a powerful but convenient method to realize the hidden chirality-induced OAM.

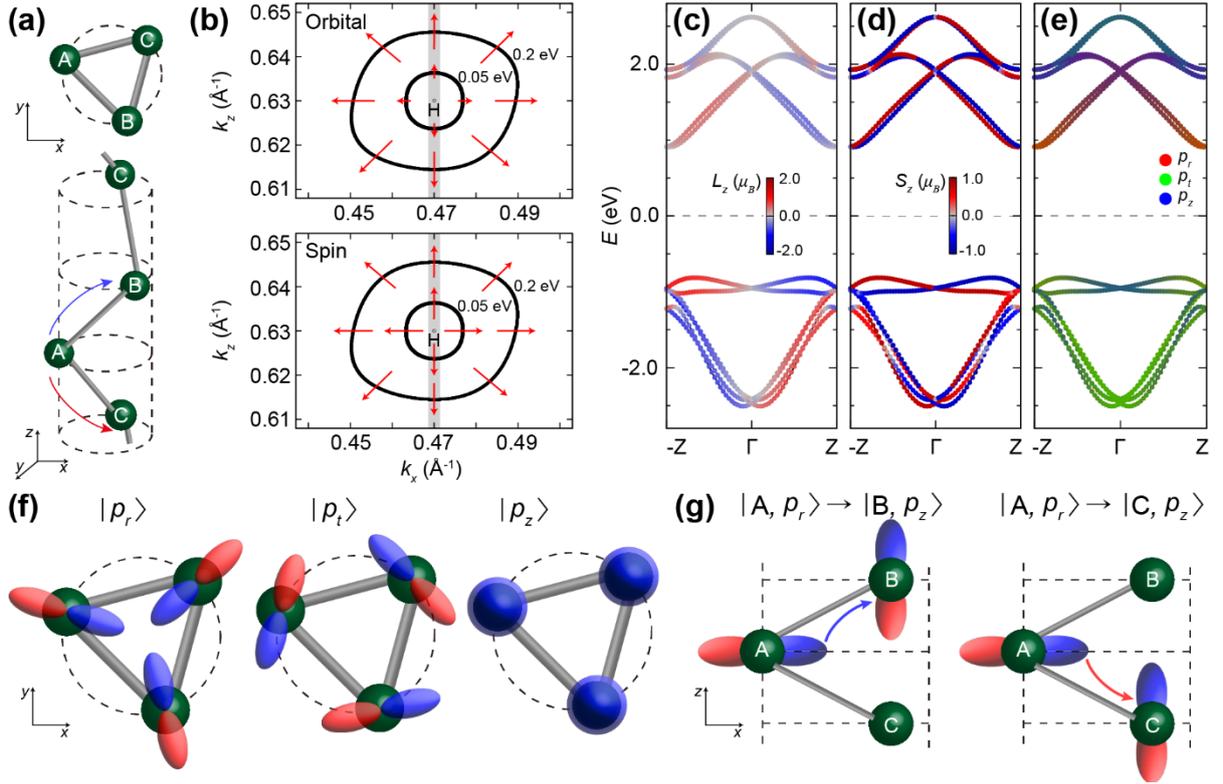

**Figure 1.** Orbital/Spin angular momentum of a helical Se chain. (a) Top and side views of a single chain of elemental Se with three different atomic sites (A, B, and C). (b) Orbital and spin angular momentum profiles of the conduction band minimum at the H point in bulk Se. Black contour lines denote energy levels of 0.05 and 0.2 eV, respectively, and gray lines represent the one-dimensional Brillouin zone of an individual Se chain. Calculated band structure of the Se chain with (c) orbital-angular-momentum resolution ($L_z$), (d) spin-angular-momentum resolution ($S_z$), and (e) atomic-orbital resolution using radial ($p_r$), tangential ($p_t$), and axial ($p_z$) p-orbitals as the orbital bases. (f) Schematic drawing of the $p_r$, $p_t$, and $p_z$ orbitals. (g) Graphical illustration of asymmetric orbital hopping between $p_r$ and $p_z$ in a single Se chain.

## Results and Discussion

**Orbital and spin angular momentum texture in a chiral Se chain**

We elucidate the correlation between the helical geometry of a chiral Se chain and its unique OAM and SAM texture through electronic structure calculations. The Se chain exhibits threefold screw-axis symmetry, as depicted in Figure 1a, and stack to form a trigonal elemental Se crystal, as shown in Figure S1. Bulk Se has a radial orbital and spin texture resembling hedgehogs near the Femi level at the H point,[21, 27] as illustrated in Figure 1b, indicating the orbital-momentum and spin-momentum locking. Interestingly, this radial OAM is preserved even without the inclusion of SOC, as shown in Figure S2 and Figure S3, while the characteristic spin structure is disrupted. Even at the A point, as illustrated in Figure S4, the radial texture disappears, but the parallel relation between $k_z$ and $L_z$ are preserved.

We investigate whether the $z$-directional orbital/spin-momentum locking of the chiral Se arises from its helical geometry of the single chain or from the staking of multiple chains. To this end, we compare the orbital and spin structure of the axial line of bulk Se (gray lines of Figure 1b) with those of the single chiral chain (Figure 1c-e). The radial angular momentum of the stack of the chains successfully corresponds to the orbital/spin polarization parallel to the chain direction ($k_z$) in the one-dimensional Brillouin zone, implying that the distinct parallel polarization can be related to the intrachain effect of the single Se chain. The absence of mirror symmetry allows for the non-vanishing out-of-plane orbital/spin component in the Se chain, and time-reversal symmetry enforces the zero net moment, consequently satisfying the relations such that $\langle \hat{L}_z \rangle_{n,k_z} = -\langle \hat{L}_z \rangle_{n,-k_z}$, and $\langle \hat{S}_z \rangle_{n,k_z} = -\langle \hat{S}_z \rangle_{n,-k_z}$. The polarized states near the Fermi level primarily originate from $p$-orbitals. To account for the helical nature of the chain, we adopt a cylindrical $p$-orbital basis. Specifically, while the valence bands consist mainly of the tangential ($p_t$) and axial ($p_z$) orbitals, the conduction bands are derived from the radial ($p_r$) orbitals (Figure 1e, f). As in the bulk, the one-dimensional orbital texture persists even in the absence of SOC as demonstrated in Figure S5. We obtained similar characteristics of OAM and SAM in a helical Te chain, as shown Figure S6.

To understand the underlying physics behind the unconventional OAM and SAM texture originating from the unique geometry of the chiral Se chain, we develop an analytic model Hamiltonian based on the tight-binding approximation.[38-41] By considering the nearest hoppings between three $p$-orbital states for each of the three atoms in the Se chain (Figure 1a), we obtain a 9 × 9 Hamiltonian matrix as follows,

$$H = \begin{pmatrix} h_{\text{on}} & h_{\text{hop}} & h_{\text{hop}}^{\dagger} \\ h_{\text{hop}}^{\dagger} & h_{\text{on}} & h_{\text{hop}} \\ h_{\text{hop}} & h_{\text{hop}}^{\dagger} & h_{\text{on}} \end{pmatrix}, \quad (1)$$

where the on-site Hamiltonian $h_{\text{on}}$ and the hopping Hamiltonian $h_{\text{hop}}$ are 3 × 3 matrices in the $p$-orbital space. The on-site Hamiltonian is modeled as $h_{\text{on}} = \epsilon_r |p_r\rangle\langle p_r| + \epsilon_t |p_t\rangle\langle p_t| + \epsilon_z |p_z\rangle\langle p_z|$, where $\epsilon_\alpha$ denotes the on-site energy for the $p_\alpha$ orbital. The use of the cylindrical basis (Figure 1f) allows us to express the off-diagonal part in terms of a single matrix $h_{\text{hop}}$, referred to as the 'hopping-down' Hamiltonian, which describes the nearest hopping in the direction of C → B → A → C. In contrast, the hopping in the opposite direction is given by $h_{\text{hop}}^{\dagger}$, the 'hopping-up' Hamiltonian. By carefully examining the signs of the hopping integrals for all hopping procedures (see Figure 1g for examples), we model $h_{\text{hop}}$ as

$$h_{\text{hop}} = e^{ik_z a} \begin{pmatrix} t_{rr} & -t_{tr} & -t_{zr} \\ t_{tr} & t_{tt} & t_{zt} \\ t_{zr} & t_{zt} & t_{zz} \end{pmatrix}, \quad (2)$$

in the ($p_r, p_t, p_z$) basis. Here $t_{\beta\alpha}$ denotes the hopping integral for $\alpha \to \beta$ hopping and $a$

represents the one-third of the length of the unit cell, which corresponds to the distance between the nearest sites along the z-axis, giving rise to the Bloch-like factor $e^{ik_z a}$. The anti-symmetric nature of $t_{tr}$ and $t_{zr}$ for the hopping-up and hopping-down, as exemplified in Figure 1g, is rooted in the unique geometry of the helical Se chain and turns out to be a crucial factor in the emergence of the OAM texture shown in Figure 1c. We demonstrate this below through equations (3) and (4).

To reproduce the OAM texture with reduced reliance on SOC (Figure S2 and Figure S5), we focus on the orbital degree of freedom by diagonalizing the Hamiltonian [equation (1)] while disregarding the spin degree of freedom. The hopping integrals are treated perturbatively to avoid the technical complexity of exact diagonalization, resulting in the nine eigenstates $|\widetilde{p_\alpha, \mu}\rangle$ presented in Note S1. In this notation, $\alpha = r, t, z$ refers to the orbital states, $\mu = -1, 0, 1$ denotes the atomic degree of freedom. The tilde on the states indicates that they are perturbed by electron hopping, and they are expressed as a superposition of the unperturbed states $|p_\beta, \mu\rangle$. In Note S2, we explicate the consistency between the eigenstates from the model Hamiltonian and the band structure from our first-principles calculations (Figure 1c-e). One can calculate the OAM expectation values from the eigenstates using $L_z = (i\hbar/2)(|p_t\rangle\langle p_r| - |p_r\rangle\langle p_t|)$ and obtain

$$\langle \widetilde{p_z, \mu}|L_z|\widetilde{p_z, \mu}\rangle = \frac{4\hbar t_{zr} t_{zt} \cos\left(k_z a + \frac{2\pi\mu}{3}\right) \sin\left(k_z a + \frac{2\pi\mu}{3}\right)}{(\epsilon_z - \epsilon_r)(\epsilon_z - \epsilon_t)}, \qquad (3)$$

$$\langle \widetilde{p_t, \mu}|L_z|\widetilde{p_t, \mu}\rangle = -\langle \widetilde{p_r, \mu}|L_z|\widetilde{p_r, \mu}\rangle = \frac{2\hbar t_{tr} \sin\left(k_z a + \frac{2\pi\mu}{3}\right)}{\epsilon_t - \epsilon_r}. \qquad (4)$$

This simplified model effectively captures the essential characteristics of the numerical results, such as the band splitting and the sign change of OAM at the zone center, as exhibited in Figure

S5. The anti-symmetric hopping enables the quantum mechanical superpositions of atomic orbitals that are energetically separated, leading to the non-vanishing orbital distribution in momentum space. In this aspect, our approach is distinct from previous models dealing with a helical geometry without incorporating the orbital degree of freedom which derive a non-zero OAM distribution from inter-atomic hoppings.[30, 31] In Figure S7, we demonstrate that the inter-atomic contribution is negligible near the Fermi level in the Se chain, and thus the consideration of the orbital degree of freedom is essential to describe the intriguing orbital structure of the helical chain.

We incorporate the spin degree of freedom into our model Hamiltonian to investigate how spin and orbital degrees of freedom are intertwined by the presence of SOC in the Se chain. In the absence of SOC, the SAM lacks any directional preference and consequently a particular spin texture does not occur. In the presence of SOC ($\eta_{SO}\mathbf{L} \cdot \mathbf{S}$), the spin degeneracy is lifted in accordance with the relative orientation of SAM and OAM [equations (3) and (4)], thereby creating a spin polarization $\langle S_z \rangle = (\nu\hbar/2)\,\text{sgn}(\langle L_z \rangle)$ that is either parallel or antiparallel to the OAM, where $\nu = \pm 1$ is the spin quantum number. The detailed treatment of physical quantities, such as orbital and spin, as well as their coupling, is presented in Note S5. Therefore, we can conclude that the OAM is a more fundamental degree of freedom that gives rise to the SAM texture in the Se chain.[42] In this regard, our model offers a concrete instance of a recent theoretical proposition,[43] which posits the essential significance of the orbital degree of freedom in the long-range spin transport in DNA-like materials.[3, 44-46]

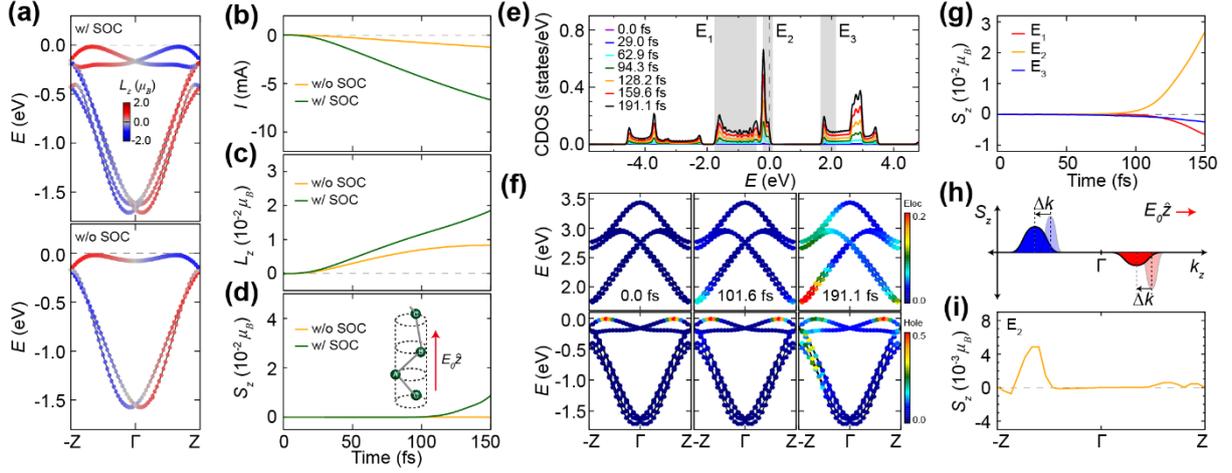

**Figure 2.** Current-induced magnetism in a helical Se chain. (a) Calculated band structure of a Se chain with orbital-angular-momentum resolution ($L_z$) under hole doping (0.1 $e$/unit) in the presence or absence of SOC. Calculated (b) induced current ($I$), (c) total orbital angular momentum ($L_z$), and (d) total spin angular momentum ($S_z$) induced by a dc field ($10^{-3}$ V/Å) along the Se chain, as illustrated in the inset of (d). The quantitative definitions of $I$, $L_z$, and $S_z$ are described in the Methods section. (e) Time evolution of the carrier density of states (CDOS), the non-equilibrium portion of excited states, projected onto the unperturbed electron/hole states in the Se chain. (f) Time evolution of the electron/hole weights projected onto the ground-state band structure of the Se chain. (g) Time evolution of the decomposed spin angular momentum of the Se chain for three energy ranges ($E_1$, $E_2$, and $E_3$) as shown in the gray regions of (e). (h) Schematic illustration of the conventional Edelstein effect, in which a net spin magnetic moment is generated by the Fermi surface shift triggered by an external field. (i) Calculated momentum-resolved spin angular momentum for the major $E_2$ contribution in the Se chain at 191.1 fs. The spin distribution of the chiral chain is distinguished from that of the conventional Edelstein effect shown in (h).

**Nonequilibrium generation of orbital and spin magnetism**

Drawing on a fundamental understanding of the interplay between the OAM and the SAM in the Se chain, we investigate the time-dependent dynamics of these quantities in non-equilibrium under the influence of an external field. By utilizing TDDFT calculations, we reproduce the excited states in a single Se chain and identify the emergent magnetism and its dominant mechanism. To simulate metallic nanowires used in the experiment, we employed a hole-doped (0.1 $e$/unit) Se chain, in which the Fermi level touches the valence bands, as shown in Figure 2a, while preserving its original OAM distribution. A detailed description of the methodology used to estimate the time-evolving electric current, OAM, and SAM is provided in the Methods section. To simulate the current flowing along the chain in the experiment,[27] we applied a gradually turn-on DC field of $10^{-3}$ V/Å (Figure S8). Unlike the immediate response of electric current and orbital magnetism, which begin to arise at 10 fs and increase monotonically (Figure 2b, c), the development of non-zero spin magnetism in the Se chain requires a substantial time delay (100 fs) and is dependent on SOC, as depicted in Figure 2d. It is known that the timescale of electronic excitation is a few femtoseconds, while the time scale of SOC is a few hundred femtoseconds.[47] These times scales align well with our calculations, where the OAM start to develop at $t \sim 10$ fs and the SAM begin to develop at $t \sim 100$ fs. This suggests that the orbital magnetism drives the spin part through SOC. Furthermore, the dominance of orbital magnetism at the early stage under an external field is also observed in a helical Te chain (Figure S9).

We conduct an in-depth analysis to understand the formation of finite orbital and spin magnetism in the Se chain by tracking the real-time nonequilibrium states projected onto the ground states. We investigate the carrier density of states for the Se chain, representing the difference in density of states between the excited and ground states. The density of states of the excited states extends over a wide range from -4 to 4 eV (Figure 2e), and we found that the overall shape is independent of the external field strength (Figure S10). Such a widespread excitation originates from the strong mixing of the orthogonal orbitals via the helical structure of the Se chain. As shown in Figure 1a, f, the shapes of the tangential orbital of the A site and the radial orbital of the B site are similar to each other, leading to considerable overlap of the orbitals under the external field along the $z$ direction. As a result, the strong mixing gives a widespread distribution in Figure 2e when projected to the unperturbed eigenstates.

The widespread density of state is observed not only in energy, but also in momentum space as shown in Figure 2f. The projection of the excited states onto the ground electronic structure exhibits an asymmetric *k*-distribution of nonequilibrium weights within the lowest conduction band near the zone boundary at 101.6 fs, while maintaining a symmetric distribution in the highest valence band. This finding indicates that the induced orbital magnetization of the early time (< 100 fs) is facilitated by the imbalanced transition to the conduction bands which have different orbital polarization (Figure 1c). The imbalanced distribution is induced by the chiral structure, which enforces the direction-dependent orbital mixing as illustrated in in Figure S11. However, the equal weights of the conduction bands with different spin polarization inhibit the occurrence of a spin magnetization (Figure 1d). Thus, the emergent total OAM remains unaffected by the absence of SOC, as demonstrated in Figure S12. At 191.1 fs, the excited state displays the imbalanced momentum distribution within the valence bands, which coincides with the onset of the spin magnetization (Figure 2f, g). The formation of spin magnetism in the chiral Se chain is primarily attributed to the partial occupancy of the highest spin-split valence bands in the $E_2$ region, as shown in Figure 2g. It is important to note that this mechanism differs from the conventional Edelstein effect, in which a shift of the Fermi surface induced by an electrostatic field give rise to a net spin polarization due to opposite spin-momentum locking at +*k* and -*k* points,[48] as illustrated in Figure 2h. In the Se chiral chain, however, the net magnetization does not originate from the spin difference between +*k* and -*k* points, but between the highest valence bands with opposite spin polarizations at the same point as shown in Figure 2i.

Our findings signify a quantum-mechanical development of orbital magnetization under an axial electric field by the imbalanced distribution of excited states, giving rise to an accompanied spin magnetization induced by the unique interconnection of orbital and spin triggered by the helical geometry. It should be noted that this study does not take into account various scattering sources, such as phonons and defects. Therefore, to avoid misinterpretation, we focus solely on the early stages of the time dynamics of the excited state.

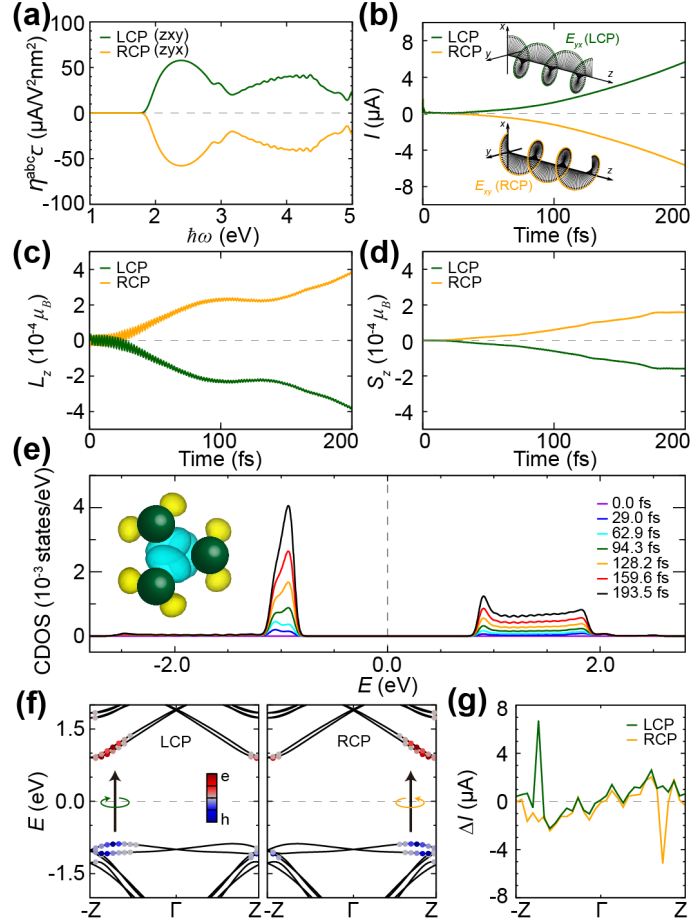

**Figure 3.** Circular photogalvanic effect in a helical Se chain. (a) Calculated circular photogalvanic current (injection current) spectrum of a Se chain as a function of the frequency of a left or right circularly polarized light (LCP or RCP). Calculated (b) induced current ($I$), (c) total orbital angular momentum ($L_z$), and (d) total spin angular momentum ($S_z$) induced by circularly polarized lights with intensity of 0.051 V/Å and frequency of $\hbar\omega$=2.02 eV as a function of time. Schematic drawings of left- and right-handed circularly polarized light (LCP and RCP) are described in the inset of Figure 3b. (e) Time evolution of the electron/hole-carrier density of states (CDOS) in the Se chain under a circularly polarized light. The inset represents a real-space representation of the generated electron/hole carriers. Yellow and cyan iso-surfaces denotes the excited hole and electron density, respectively. (f) Electron/hole carrier occupancies projected onto the ground-state band structure of the Se chain under LCP or RCP at $t$=193.4 fs. (g) Momentum-resolved photogalvanic current density under LCP or RCP in the

Se chain at t=193.4 fs.

**Optoelectronic detection of orbital angular momentum**

In order to appreciate the physical significance of the OAM generated in the chiral structure, we shall proceed to examine a comprehensive operator expression of the OAM. The lateral confinement of electron motion to a finite region in a one-dimensional system enables an expression of the $L_z$ operator in terms of position operators : $L_z = (m_e/2i\hbar)(\{x,[y,H]\} - \{y,[x,H]\})$, where $m_e$ is the electron mass and $H$ is the Hamiltonian. The OAM expectation value for an eigenstate $|n\rangle$ of $H$ is then $\langle n|L_z|n\rangle = (2m_e/\hbar)\text{Im}\sum_{n'}(\epsilon_n - \epsilon_{n'})\langle n|x|n'\rangle\langle n'|y|n\rangle$, where $\epsilon_n$ is the energy eigenvalue of $|n\rangle$. Provided that the dominant contribution of the OAM comes from the pair of the highest valence band (*v*) and the lowest conduction band (*c*), the OAM is simply given by

$$\langle c|L_z|c\rangle \approx \langle v|L_z|v\rangle \approx \frac{2m_e}{\hbar}\text{Im}[(\epsilon_c - \epsilon_v)\langle c|x|v\rangle\langle v|y|c\rangle]. \tag{5}$$

The right-hand side is referred as to the two-band OAM, denoted as $\langle L_z\rangle_{cv}$. In Figure S13, we offer numerical validation of the two-band approximation in our system by demonstrating the resemblance between the total OAM and the two-band OAM. The significance of the two-band OAM in the reduced dimension lies in its feasibility to be measured by the circular photogalvanic effect (CPGE) beyond mere theoretical speculation. By modifying the well-established CPGE formalism with our two-band OAM,[49] we obtain the circular photogalvanic current $J$ (i.e., injection current) in the low-dimension. Specifically, the injection current is given by

$$J = -\frac{2\pi\tau}{m_e\hbar^2\omega}e^3E^2\langle L_z\rangle_{cv}, \tag{6}$$

where $\omega$ and $E$ is the frequency and the electric field of the incident light, respectively, and $\tau$ is the relaxation time. The derivation of equation (6) is explicated in Note S3. In two or three dimensions with periodic boundary conditions, the CPGE is explained in terms of two-band Berry curvature.[49] However, in a one-dimensional system, the Berry curvature is ill-defined and requires a generalization through the adoption of position operators.[50] Conversely, the OAM is inherently defined in a confined space, making it a more suitable physical quantity to quantify the optical response of a low-dimensional system.

To explicitly demonstrate the intimate relationship between the OAM and the CPGE, we utilize second-order perturbation theory to evaluate the response function of the injection current. As predicted in equation (6), the injection current is created under a circularly polarized light with above-band-gap frequency in the helical Se chain (Figure 3a). The injection current can be switched by the chirality of the chain as well as the helicity of light (Figure S14), indicating the generated photocurrent serves as a reference for determining the chain's chirality. The detailed optical properties, such as the linear absorption spectrum and circular dichroism,[51] are illustrated in Figure S15.

We also confirm the photocurrent generation in the Se chain by applying a circularly polarized light ($\hbar\omega$=2.02 eV) through TDDFT calculations as presented in Figure 3b. As a secondary effect owing to the selective excitation of orbital/spin-polarized electrons, the generated current carries induced orbital and spin magnetism in the chain (Figure 3c, d), which can be controlled by the helicity of applied light even in a very strong field limit (Figure S16). We also observed a similar relationship between the OAM and CPGE in the Te chain (Figure S17). The monotonic increasing number of electron and hole carriers near the Fermi level with time (Figure 3e) clearly implies that the CPGE of the Se chain is governed by the band edge transition corresponding to the frequency of the applied light. As expected from Figure 1e, the hole (electron) carriers are predominantly composed of the $p_t$ ($p_r$) orbitals in the CPGE of the Se chain (the inset of Figure 3e). The projection of the excited state onto the ground band structure exhibits uneven excitation caused by circularly polarized light (Figure 3f),[52, 53] leading to a non-zero photocurrent that strongly localizes near the zone boundary and depends on the helicity of external light (Figure 3g). Hence, the intriguing OAM texture embedded in the chiral structure can be readily realized by detecting the generated photocurrent in response to circularly polarized light.

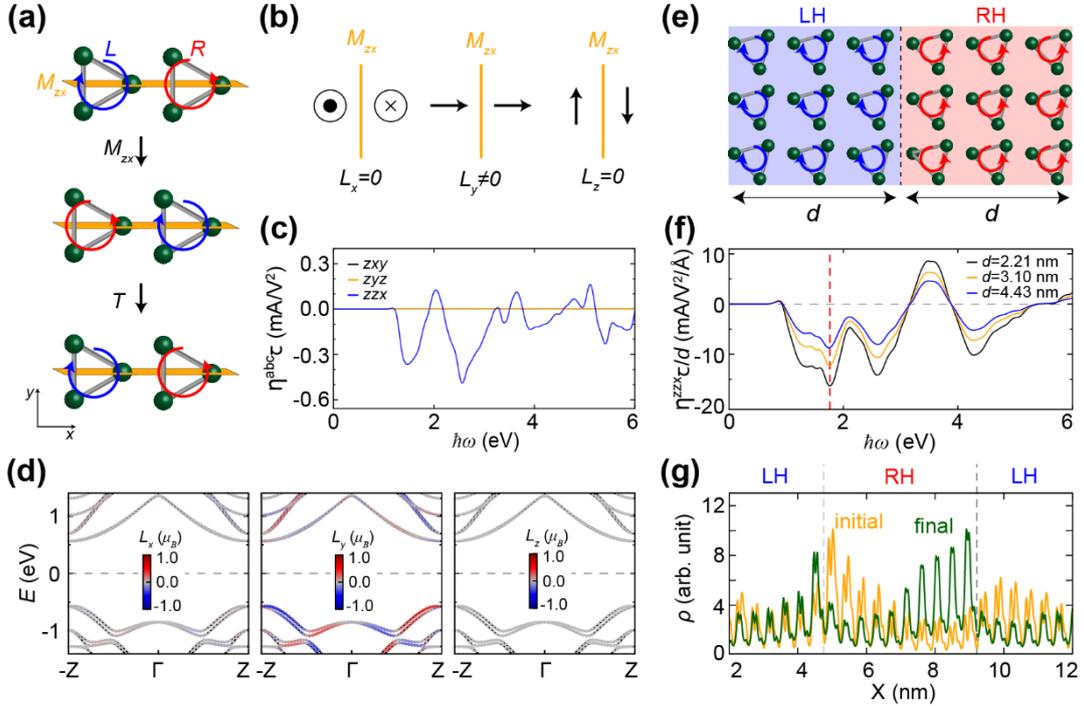

**Figure 4.** Detection of boundary states using the circular photogalvanic effect. (a) Schematic drawing of hypothetical two-dimensional helical Se layers constating of alternating left-handed (*L*) and right-handed (*R*) configurations. The alternating chains are invariant under consecutive *zx* mirror ($M_{zx}$) and translation (*T*) operations. (b) $M_{zx}$ mirror operation on each component of the orbital angular momentum in the alternating chains, which enforces the vanishing *x*-and *z*-components of orbital angular momentum along the one-dimensional chain direction. (c) Calculated circular photogalvanic current (injection current) spectrum of the alternating chains as a function of the frequency of applied circularly polarized lights. The *zx*-plane polarization (blue) induces a non-zero photocurrent, while *xy*-plane (black) or *yz*-plane (yellow) polarization result in a zero current. (d) Calculated

band structure of the alternating chains with orbital-angular-momentum resolution. (e) Schematic drawing of a boundary between left-handed (LH) and right-handed (RH) chains.(f) Variation in the injection current spectrum per unit length of the domain structure upon the domain length ($d$). (g) The average charge density profiles of the initial and final states of an excitation triggered by an external light with $\hbar\omega$ = 1.76 eV, corresponding to the red dotted line in (f).

We extend the concept of optoelectronic detection of the concealed orbital polarization to a more intricate and realistic system, wherein extrinsic OAM that is otherwise prohibited in bulk is enforced by spatial symmetry. As a simplified case study to elucidate the relationship between the presence of photocurrent and OAM determined by geometric symmetry, we consider a two-dimensional Se structure with alternating chirality along the *x* direction as illustrated in Figure 4a. Our hypothetical structure possesses a combined symmetry of *zx*-plane mirror ($M_{zx}$) and translation (*T*), which nullifies the *x* and *z* components of OAM as shown in Figure 4b (see Note S4 for a more comprehensive elucidation). One can easily generalize the photogalvanic current of equation (6) for different light proportion directions, and the survival of the $L_y$ component can be interpreted as the activation of photocurrent under circularly polarized light with a *zx* rotating plane. In line with our symmetry analysis, the alternating two-dimensional system allow for the presence of the *y* component of OAM (Figure 4d), which is absent in a single Se chain (Figure S18), leading to a non-zero photocurrent in a wide energy range under *zx*-circularly polarized light (Figure 4c).

The essential physics underlying the artificial alternating chains can be applied to a more pragmatic system featuring opposite chiralities, namely the domain boundary between left-handed (LH) and right-handed (RH) domains as presented in Figure 4e. The circular photogalvanic current along the chain under the *zx*-circular polarization, normalized by the domain size (*d*), decreases as the domain size increases (Figure 4f), implying its interface-localized nature resulting from the switching of the chirality. We check that excited carriers generated by an external light with $\hbar\omega$ = 1.76 eV (red vertical line in Figure 4f) are distributed across the domain boundary by computing the charge density distribution of the initial and final states (Figure 4g). Importantly, this interface photocurrent is not observed in the bulk without a boundary (Figure S19). In the case of an infinite homochiral layer without a domain boundary, the rotation of the system around the $z$ axis by 180º results in a sign change of $E_x$ while preserving the chirality of each of the chains. This transformation implies that $\eta_{zzx}$ should be transformed to -$\eta_{zzx}$, leading to its vanishing. The strong correlation between OAM and photocurrent is further evidenced by the non-negligible $L_y$ value observed across the entire $k_z$ line (Figure S20). The states with a non-zero $L_y$ component are localized at the chiral boundary between the left-handed and right-handed Se chains. (Figure S21). The chiral boundary photocurrent is observed to be robust even at the boundary of Te, as depicted in Figure S22. Such a photocurrent is also produced at the edge between Se chains and vacuum as shown in Figure S23, signaling the viability and applicability of OAM engineering and its optoelectronic detection.

**Conclusions**

In summary, we revealed the quantum mechanical origin of the chirality-driven OAM and its manifestation by means of electrical/optical drives in a helical Se chain and narrow down the discrepancy between theory and experiment. Our model for the helical chain showed the anti-symmetric chiral hopping produces a unique orbital texture parallel to its momentum, generating a radial spin profile with the aid of strong SOC. The intermixing of orthogonal *p*-orbital states and asymmetric nonequilibrium distribution gives rise to a net orbital magnetism under an external electric field, accompanied by a net spin magnetism due to the partial occupancy of spin-split states in the presence of SOC. This induced magnetism is quantum mechanically distinguished from the conventional Edelstein effect in Rashba SOC systems. We also demonstrated the generation of photocurrent can be expressed in terms of OAM in low dimensions, which provides an alternative to the poorly defined Berry curvature in a non-periodic system. The close connection between OAM and photocurrent enables the optoelectronic manifestation of the intriguing orbital polarization inherent in the chiral chain. Furthermore, we showed that, by tuning the geometric symmetry at an interface or edge, a previously forbidden component of OAM can be allowed, leading to the generation of edge-localized photocurrent. Our results provide a fundamental understanding of the interplay between spin and orbital degrees of freedom in chiral geometries, which would be beneficial for design of next generation orbitronic/spintronic devices utilizing chiral materials.

## Methods

**Electronic structure calculation.** We carried out density functional theory (DFT) calculations using the projector-augmented plane-wave method,[54] as implemented in the Vienna Ab initio Simulation Package (VASP).[55] The Perdew–Burke–Ernzerhof (PBE) functional of the generalized gradient approximation was employed to describe the exchange–correlation interactions among electrons.[56] The bulk Se chains were optimized with the van der Waals interaction until the maximum forces were less than 0.001 eV Å$^{-1}$, and the vacuum layer was set to be greater than 10 Å to simulate the one-dimensional Se chain. The energy cutoff for the plane-wave-basis expansion was set at 400 eV. We used a 10×10×8 $k$-point grid for the bulk Se chains and a 1×1×15 $k$-point grid for the one-dimensional Se chain.

**Circular photogalvanic response using perturbation theory.** The injection-current spectra were evaluated from the tight-binding Hamiltonian based on maximally localized Wannier functions using the second-order optical response formalism.[57, 58]

$$\eta^{abc}(\omega) = \frac{e^3 \pi}{2\hbar^2} \int \frac{d\mathbf{k}}{8\pi^3} \sum_{nm} \Delta_{mn}^a f_{nm} (r_{mn}^c r_{nm}^b - r_{mn}^b r_{nm}^c) \delta(\omega_{mn} - \omega)$$

where $a$, $b$, $c$ are Cartesian indices and $r_{nm}^a = i\langle n|\partial_{k_a}|m\rangle$ is interband Berry connection. $f_{nm}$ is occupation difference between band $n$ and $m$, and $\hbar\Delta_{mn}^a = v_{mm}^a - v_{nn}^a$ is the group velocity difference between band $n$ and $m$. For the injection-current estimation, a 3×3×100 $k$-point grid was adopted for calculating the one-dimensional Se chains. This is a sufficiently dense mesh of grids to show convergence of the injection-current spectra.

**Real-time calculation of the electric current, OAM, and SAM.** To examine the electric current, OAM, and SAM, we performed real-time TDDFT calculations using the plane-wave-based real-time evolution.[59, 60] In our calculations, the Kohn–Sham wavefunction, the density, and the Hamiltonian were self-consistently evolved through the time-dependent equation:

$$i\hbar \frac{\partial}{\partial t} \psi_{n,\mathbf{k}}(\mathbf{r},t) = \left[ \frac{1}{2m_e}\left(\frac{\hbar}{i}\nabla + \frac{e}{c}\mathbf{A}_{\text{ext}}(t)\right)^2 + \sum_\lambda v_{\text{atom}}(\mathbf{R}_\lambda(t)) + V_{\text{DFT}}[\rho(\mathbf{r},t)] \right] \psi_{n,\mathbf{k}}(\mathbf{r},t)$$

where $n$ and $\mathbf{k}$ denote the band index and the Bloch momentum vector, respectively. $\mathbf{A}_{\text{ext}}$ and $V_{\text{DFT}}$ indicate the time-dependent vector potential and DFT potential, respectively. The discretized time step for the time integration ($\Delta t$) was set to 2.415 as. In our calculations, the electric field was expressed using the velocity gauge of the vector potential via the relation $\mathbf{E}(t) = -\frac{1}{c}\partial \mathbf{A}_{\text{ext}}/\partial t$. We considered two different electric fields in our calculation: DC-field $\mathbf{E}_{\text{DC}}(t) = E_{\text{DC}}\hat{z}$ and AC-field $\mathbf{E}_{\text{AC}}(t) = E_{\text{AC}}\sin(\omega t)\hat{z}$. The initial wavefunctions [$\psi_{n,\mathbf{k}}(t=0)$] were obtained from the static-ground-state DFT calculations using the QUANTUM ESPRESSO package with the PBE exchange–correlation functional. Using the time-evolving Bloch wavefunctions, we evaluated the time profile of the electric current, OAM, and SAM as follows:[61, 62]

$$I(t) = -\frac{e}{m_e} \sum_n \sum_\mathbf{k} f_{n,\mathbf{k}} \langle \psi_{n,\mathbf{k}}(t) | \hat{\boldsymbol{\pi}} | \psi_{n,\mathbf{k}}(t) \rangle$$

$$L_z(t) = \sum_n \sum_\mathbf{k} f_{n,\mathbf{k}} \langle \psi_{n,\mathbf{k}}(t) | \hat{x}\hat{\pi}_y - \hat{y}\hat{\pi}_x | \psi_{n,\mathbf{k}}(t) \rangle$$

$$S_z(t) = \sum_n \sum_\mathbf{k} f_{n,\mathbf{k}} \langle \psi_{n,\mathbf{k}}(t) | \hat{S}_z | \psi_{n,\mathbf{k}}(t) \rangle$$

where $n$ is the band index, $f_{n,\mathbf{k}}$ is the initial occupation of the Bloch state, $m_e$ is the mass of an electron. The gauge-invariant mechanical momentum is defined as $\hat{\boldsymbol{\pi}} = \frac{m_e}{i\hbar}[\hat{\mathbf{r}}, \hat{H}] = \hat{\mathbf{p}} + \frac{e}{c}\mathbf{A}_{\text{ext}}(t) + i\frac{m_e}{\hbar}[V_{\text{NL}}, \hat{\mathbf{r}}]$. Here $\hat{x}, \hat{y}$ are the position operator along the $x$- and $y$-axis, respectively, and $\hat{S}_z$ is the SAM operator along the $z$-axis. To estimate the carrier density of states (CDOS) in Figure 2e and the nonequilibrium carrier weight ($W_{n,k}$) in Figure 2g, we projected the time-evolving wavefunction onto the unperturbed ground states as below[62, 63];

$$\text{CDOS}(E) = \left| \sum_{m,\mathbf{k}} \frac{1}{\sigma\sqrt{2\pi}} \left[ 1 - \sum_n f_{n,\mathbf{k}} |\langle \psi_{n,\mathbf{k}}(t) | \psi_{m,\mathbf{k}}(t=0) \rangle|^2 \right] e^{-\frac{1}{2}\frac{(E-\epsilon_{m,\mathbf{k}})^2}{\sigma^2}} \right|,$$

$$W_{n,\mathbf{k}}(t) = \sum_m \left[ f_{m,\mathbf{k}} |\langle \psi_{n,\mathbf{k}}(t) | \psi_{m,\mathbf{k}}(t=0) \rangle|^2 \right] - f_{n,\mathbf{k}},$$

where $\sigma$ is the smearing parameter, $n$ and $m$ are band indices, and $k$ is the one-dimensional $k$-grid. $f_{n,k}$ represents the occupation number at specific Bloch states. $\psi_{m,k}(t=0)$ is the unperturbed states while $\psi_{n,k}(t)$ represents time-evolving wavefunction. For treating the relativistic effect, we exclusively considered spin-orbit coupling[$\sim e\boldsymbol{\sigma} \cdot (\mathbf{A} \times \nabla V)$] and omitted its light-induced field correction [$\sim e\boldsymbol{\sigma} \cdot (\mathbf{A} \times \nabla V)$], as the contribution off the light-induced field correction (0.71×10⁻⁷ eV) is negligible compared to the spin-orbit coupling contribution (0.13 eV).

## ASSOCIATED CONTENT

A preprint version of this manuscript has been previously submitted to the arXiv preprint server accessible at https://arxiv.org/abs/2304.09364.[64]

**Supporting Information**

The Supporting Information is available free of charge at xxxx.

- The Supporting Information PDF contains the following sub-sections: Diagonalization of the model Hamiltonian for a one-dimensional chiral Se chain (Note S1), Band characterization of the analytic eigenstates (Note S2), The relation between orbital angular momentum and injection current in a one-dimensional chiral Se chain (Note S3), Symmetry analysis of the emergence of nonzero orbital angular momentum components in bulk Se (Note S4), justification for the spin texture derived from the orbital texture (Note S5), and supporting Figures S1-S23

## AUTHOR INFORMATION

**Corresponding Author**

Kyoung-Whan Kim - Center for Spintronics, Korea Institute of Science and Technology, Seoul 02792, Korea; https://orcid.org/0000-0002-1382-7088; Email: kwk@kist.re.kr

Jeongwoo Kim - Department of Physics, Incheon National University, Incheon 22012, Korea; https://orcid.org/0000-0002-4070-1878; Email: kjwlou@inu.ac.kr

**Author**

Bumseop Kim - Graduate School of Semiconductor Materials and Devices Engineering, Ulsan National Institute of Science and Technology, Ulsan 44919, Korea; Department of Physics, Ulsan National Institute of Science and Technology, Ulsan 44919, Korea

Dongbin Shin - Department of Physics and Photon Science, Gwangju Institute of Science and Technology, Gwangju 61005, Korea; Max Planck Institute for the Structure and Dynamics of

Matter and Center for Free Electron Laser Science, Hamburg 22761, Germany; 0000-0001-8073-2954

Seon Namgung - Department of Physics, Ulsan National Institute of Science and Technology, Ulsan 44919, Korea

Noejung Park - Graduate School of Semiconductor Materials and Devices Engineering, Ulsan National Institute of Science and Technology, Ulsan 44919, Korea; Department of Physics, Ulsan National Institute of Science and Technology, Ulsan 44919, Korea; 0000-0002-2359-0635

**Author contributions**

These authors contributed equally: Bumseop Kim, Dongbin Shin

**Notes**

The authors declare no competing interests

# Acknowledgements

We thank Binghai Yan for the initial motivation and helpful discussion. B.K. and N.P. were supported by the National Research Foundation of Korea (NRF) grant funded by the Korea government (MSIT) (No. RS-2023-00257666, No. RS-2023-00208825, and RS-2023-00218799) and were supported by Samsung Electronics Co. through Industry-University Cooperation Project (IO221012-02835-01). D.S. was supported by the National Research Foundation of Korea (NRF) grant funded by the Korea government (MSIT) (No. RS-2023-00241630). J.K. was supported by an NRF grant funded by the Korea government (MSIT) (No. NRF-2022R1F1A1059616). K.-W.K. was supported by the KIST institutional programs (2E32251, 2E32252).

manifestation of orbital angular momentum driven by chiral hopping in helical Se chains. *arXiv* (*Mesoscale and Nanoscale Physics*), **2023**, 2304.09364, ver. 2. https://arxiv.org/abs/2304.09364 (accessed May 2, 2023)

# Supporting Information for

# Optoelectronic manifestation of orbital angular momentum driven by chiral hopping in helical Se chains

Bumseop Kim[1,2†], Dongbin Shin[3,4†], Seon Namgung[2], Noejung Park[1,2], Kyoung-Whan Kim[5*], Jeongwoo Kim[6*]

[1]Graduate School of Semiconductor Materials and Devices Engineering, Ulsan National Institute of Science and Technology, Ulsan 44919, Korea

[2]Department of Physics, Ulsan National Institute of Science and Technology, Ulsan 44919, Korea

[3]Department of Physics and Photon Science, Gwangju Institute of Science and Technology, Gwangju 61005, Korea

[4]Max Planck Institute for the Structure and Dynamics of Matter and Center for Free Electron Laser Science, Hamburg 22761, Germany

[5]Center for Spintronics, Korea Institute of Science and Technology, Seoul 02792, Korea

[6]Department of Physics, Incheon National University, Incheon 22012, Korea

**Note S1. Diagonalization of the model Hamiltonian for a one-dimensional chiral Se chain.**

Here we diagonalize the 9 × 9 matrix

$$H = \begin{pmatrix} h_{\text{on}} & h_{\text{hop}} & h_{\text{hop}}^\dagger \\ h_{\text{hop}}^\dagger & h_{\text{on}} & h_{\text{hop}} \\ h_{\text{hop}} & h_{\text{hop}}^\dagger & h_{\text{on}} \end{pmatrix}, \tag{S1}$$

in the $|A\rangle, |B\rangle, |C\rangle$ atomic basis. Here $h_{\text{on}}$ and $h_{\text{hop}}$ are given by the following 3 × 3 matrices, respectively.

$$h_{\text{on}} = \begin{pmatrix} \epsilon_r & 0 & 0 \\ 0 & \epsilon_t & 0 \\ 0 & 0 & \epsilon_z \end{pmatrix}, \quad h_{\text{hop}} = e^{ik_z a}\begin{pmatrix} t_{rr} & -t_{tr} & -t_{zr} \\ t_{tr} & t_{tt} & t_{zt} \\ t_{zr} & t_{zt} & t_{zz} \end{pmatrix}, \tag{S2}$$

in the $|p_r\rangle, |p_t\rangle, |p_z\rangle$ orbital basis. We first diagonalize the atomic degree of freedom. One can show that equation (S1) is diagonalized by the following eigenstates.

$$|\mu\rangle = \frac{1}{\sqrt{3}}\left(e^{-2\pi i\mu/3}|A\rangle + |B\rangle + e^{2\pi i\mu/3}|C\rangle\right), \tag{S3}$$

where $\mu = -1, 0, 1$. The corresponding eigen-Hamiltonians are given by

$$h_{\text{on}} + 2\text{Re}\left[e^{2\pi i\mu/3} h_{\text{hop}}\right] \tag{S4}$$

which is 3 × 3 to be diagonalized. Here the real part of a matrix is defined by $\text{Re}[X] = (X + X^\dagger)/2$. More explicitly, equation (S4) is given by

$$\begin{pmatrix} \epsilon_r + 2t_{rr}\cos\tilde{k}_{z,\mu}a & -2it_{tr}\sin\tilde{k}_{z,\mu}a & -2it_{zr}\sin\tilde{k}_{z,\mu}a \\ 2it_{tr}\sin\tilde{k}_{z,\mu}a & \epsilon_t + 2t_{tt}\cos\tilde{k}_{z,\mu}a & 2t_{zt}\cos\tilde{k}_{z,\mu}a \\ 2it_{zr}\sin\tilde{k}_{z,\mu}a & 2t_{zt}\cos\tilde{k}_{z,\mu}a & \epsilon_z + 2t_{zz}\cos\tilde{k}_{z,\mu}a \end{pmatrix}, \tag{S5}$$

where $\tilde{k}_{z,\mu}a = k_z a + 2\pi\mu/3$. Since $|k_z a| \leq \pi/3$, $\tilde{k}_{z,\mu}$ satisfies $|\tilde{k}_{z,\mu}a| \leq \pi/a$. Now the physical meaning of $\tilde{k}_{z,\mu}$ is clear: a generalized Bloch momentum in a quasi-extended Brillouin zone. Ignoring the helical rotation, the three atoms in the unit cell are located periodically so that the Brillouin zone can be extended to a three times larger space. We call this a quasi-extended Brillouin zone since the helical rotation makes such a direct extension impossible. The cost of the extension is the introduction of off-diagonal hopping terms in equation (S5), which corresponds to a discretized version of the gauge potential.

Since exact diagonalization of equation (S5) is difficult, we treat $t_{\beta\alpha}$ perturbatively. Perturbation theory gives the following eigenstates of equation (5).

$$|\widetilde{p_r}\rangle = |p_r\rangle + \frac{2it_{tr}\sin \tilde{k}_{z,\mu}a}{\epsilon_r - \epsilon_t}|p_t\rangle + \frac{2it_{zr}\sin \tilde{k}_{z,\mu}a}{\epsilon_r - \epsilon_z}|p_z\rangle, \tag{S6}$$

$$|\widetilde{p_t}\rangle = |p_t\rangle - \frac{2it_{tr}\sin \tilde{k}_{z,\mu}a}{\epsilon_t - \epsilon_r}|p_r\rangle + \frac{2t_{zt}\cos \tilde{k}_{z,\mu}a}{\epsilon_t - \epsilon_z}|p_z\rangle, \tag{S7}$$

$$|\widetilde{p_z}\rangle = |p_z\rangle - \frac{2it_{zr}\sin \tilde{k}_{z,\mu}a}{\epsilon_z - \epsilon_r}|p_r\rangle + \frac{2t_{zt}\cos \tilde{k}_{z,\mu}a}{\epsilon_z - \epsilon_t}|p_t\rangle. \tag{S8}$$

From the expressions equation (S3) and (S6)-(S8), the eigenstates of the full Hamiltonian [equation (S1)] are then given by

$$|\widetilde{p_r,\mu}\rangle = |\widetilde{p_r}\rangle \otimes |\mu\rangle, \tag{S9}$$

where $\otimes$ is the direct tensor product of the two spaces.

**Note S2. Band characterization of the analytic eigenstates.**

To match the eigenstates [equations (S6)-(S8)] with the results from the first-principles calculations (Supplementary Figure 3), we note that $|\widetilde{p_\alpha}\rangle$ is the perturbed eigenstate due to the hopping elements. Therefore, without hopping, *i.e.*, $\tilde{k}_{z,\mu} = 0$ gives the unperturbed eigenstates. Therefore, the orbital character at the zone center gives the characterization of each eigenstates. For instance, the top band in Supplementary Fig. 3b is $|\widetilde{p_z}\rangle$. Since $\tilde{k}_{z,\mu}$ is a momentum in a quasi-extended Brillouin zone, there are corresponding three bands in the first Brillouin zone. Therefore, it is natural to conclude that the top three bands correspond to $|\widetilde{p_z,\mu}\rangle$ for three $\mu$ values. Similarly, the three valence bands correspond to $|\widetilde{p_t,\mu}\rangle$. Our calculation for a wider energy range (not shown) shows that $|\widetilde{p_r,\mu}\rangle$ are located far below $|\widetilde{p_t,\mu}\rangle$.

**Note S3. The relation between orbital angular momentum and injection current in a one-dimensional chiral Se chain.**

According to Ref. 48 in main text, the photogalvanic current induced by a circularly-polarised light is given by

$$J = -\frac{4\pi e \tau}{\hbar} \int \frac{dk_z}{2\pi} [\Delta v_z(k_z)] \nu(k_z) \delta(\Delta\epsilon(k_z) - \hbar\omega), \quad (S10)$$

where $\tau$ is the relaxation time, $v_z(k_z)$ is the group velocity, $\omega$ is the frequency of the incident light, $\Delta$ means the difference between the initial and final states, and $\nu(k_z) = |P^{RCP}(k_z)|^2 - |P^{LCP}(k_z)|^2$ is the optical absorption efficiency difference for the right and left circular polarised lights. Using $\hbar v_z(k_z) = \partial_{k_z}\epsilon$, equation (S10) is converted to an energy integration.

$$J = -\frac{4\pi e \tau}{\hbar^2} \int \frac{d[\Delta\epsilon(k_z)]}{2\pi} \nu(k_z) \delta(\Delta\epsilon(k_z) - \hbar\omega) = -\frac{4\pi e \tau}{\hbar^2} \sum_{\Delta\epsilon(k_z)=\hbar\omega} \nu(k_z). \quad (S11)$$

When we consider a particular $k_z$ value satisfying $\Delta\epsilon(k_z) = \hbar\omega$, we do not need to keep the summation.

To calculate $P^{RCP/LCP}(k_z)$, it is necessary to calculate the matrix element of the perturbing electric field of the light. In Ref. 48, the system is periodic so that the velocity gauge was used, *i.e.*, $P^{RCP/LCP} = (eA/m_e)\langle v|(p_x \pm ip_y)|c\rangle$, where $A$ is the magnitude of the incident vector potential. However, for our case, the system is confined along the $x$ and $y$ direction so that the position gauge is more preferred, *i.e.*, $P^{RCP/LCP} = eE\langle v|(x \pm iy)|c\rangle$ where $E$ is the magnitude of the incident electric field. Then, one can show that $\nu = e^2E^2 \text{Im}[\langle I|x|F\rangle\langle F|y|I\rangle]$, where $I$ and $F$ are the initial and the final states, respectively. As a result, we obtain

$$J = -\frac{4\pi e \tau}{\hbar^2} e^2 E^2 \text{Im}[\langle c|x|v\rangle\langle v|y|c\rangle], \quad (S12)$$

and if there are multiple $k_z$ points satisfying $\Delta\epsilon(k_z) = \hbar\omega$, the contributions should be summed up. Combining equation (S12) and equation (5) gives equation (6) in the main text.

**Note S4. Symmetry analysis of the emergence of nonzero orbital angular momentum components in bulk Se.**

We consider a general three-dimensional system with time-reversal symmetry and a mirror symmetry with respect to the *zx* plane. First, the time-reversal symmetry implies that all the physical relations should be invariant under the transformation $\mathbf{k} \to -\mathbf{k}$ and $\mathbf{L} \to -\mathbf{L}$. In other words,

$$L_x(k_x, k_y, k_z) = -L_x(-k_x, -k_y, -k_z),$$
$$L_y(k_x, k_y, k_z) = -L_y(-k_x, -k_y, -k_z), \quad \text{(S13)}$$
$$L_z(k_x, k_y, k_z) = -L_z(-k_x, -k_y, -k_z).$$

Similarly, the *zx* mirror symmetry implies that all the physical relations should be invariant under the transformation $(k_x, k_y, k_z) \to (k_x, -k_y, k_z)$ and $(L_x, L_y, L_z) \to (-L_x, L_y, -L_z)$. In other words,

$$L_x(k_x, k_y, k_z) = -L_x(k_x, -k_y, k_z),$$
$$L_y(k_x, k_y, k_z) = L_y(k_x, -k_y, k_z), \quad \text{(S14)}$$
$$L_z(k_x, k_y, k_z) = -L_z(k_x, -k_y, k_z).$$

Now we consider a more simplified case that the system is confined along the *x* and *y* directions. Then, $k_x$ and $k_y$ become meaningless. By the same argument as equations (S13)-(S14), dropping $k_x$ and $k_y$, one obtains

$$L_x(k_z) = -L_x(k_z),$$
$$L_y(k_z) = L_y(k_z), \quad \text{(S15)}$$
$$L_z(k_z) = -L_z(k_z).$$

Therefore, $L_x(k_z) = L_z(k_z) = 0$ and only $L_y(k_z)$ can survive, which is depicted in Fig. 4b.

**Note S5. Justification for the spin texture derived from the orbital texture.**

When the total Hamiltonian is expressed as the sum of the orbital part ($H_L$), the spin part ($H_S$), and the spin-orbit coupling (SOC, $H_{L\text{-}S}$), *i.e.*, $H = H_L + H_S + H_{L\text{-}S}$, the inclusion of the SOC term necessitates the re-diagonalization of the Hamiltonian. However, it is important to note that our argument simplifies the situation under specific conditions: (i) when $H_S$ is negligible, and (ii) when $H_{L\text{-}S}$ is not sufficiently large to favor the $J = L+S$ basis over the $L$-$S$ basis. The validity of the first condition is supported by a separate DFT calculation conducted without the inclusion of SOC, which demonstrates the absence of a spin texture while confirming the presence of a distinct orbital texture. The second condition is fulfilled by two supportive results, the separate behaviors of the OAM and the SAM in Figure 2c and Figure 2d, and the well-separated orbital projection in Figure 1e.

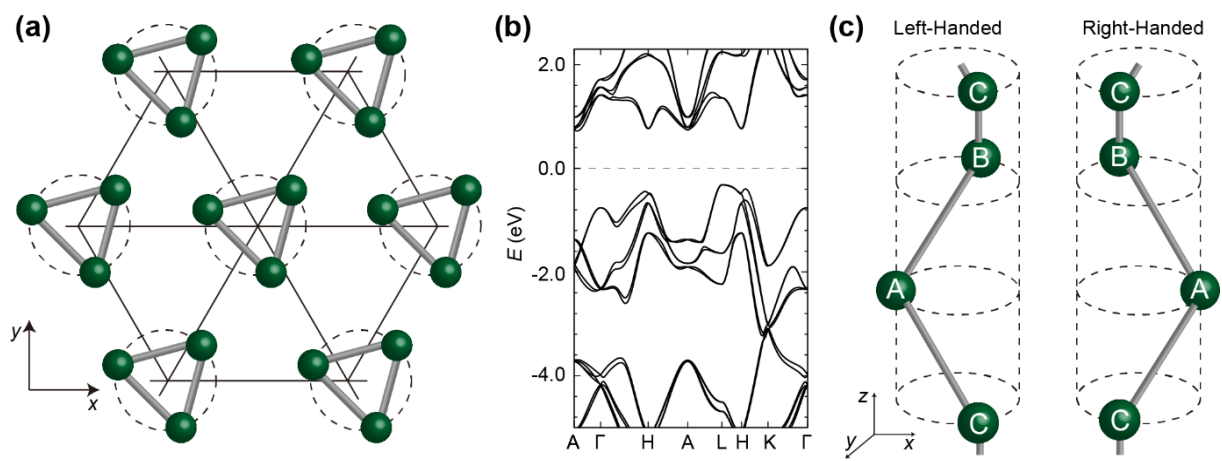

**Figure S1.** (a) Atomic and (b) electronic structure of bulk Se. (c) Schematic illustrations depicting the left-handed and right-handed crystal orientations.

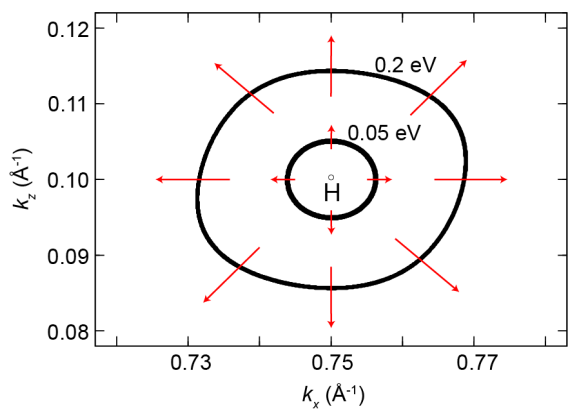

**Figure S2.** Orbital angular momentum profiles of the conduction band minimum at the H point in bulk Se without the inclusion of SOC. Black contour lines denote energy levels of 0.05 and 0.2 eV, respectively.

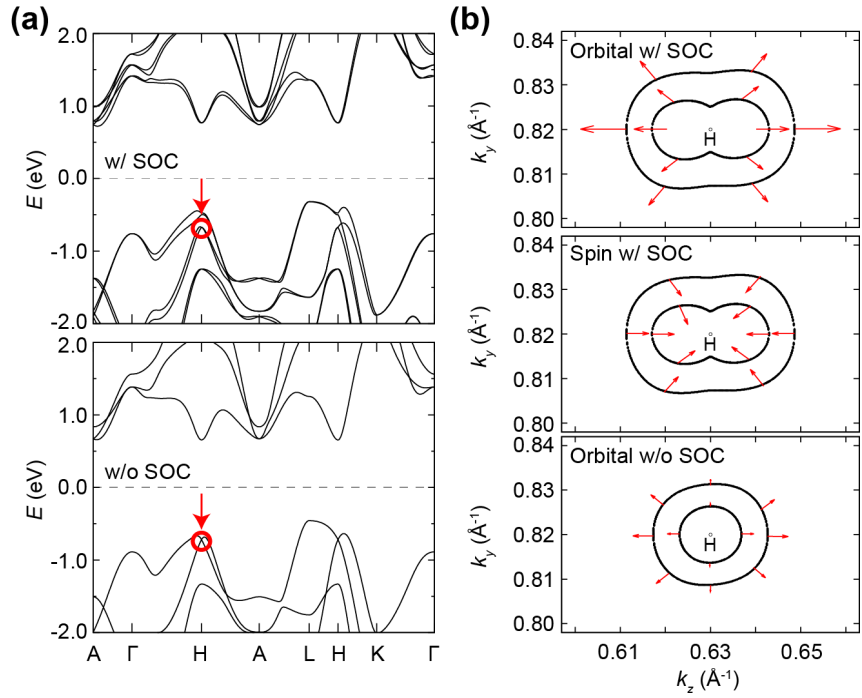

**Figure S3.** Orbital/Spin angular momentum of a helical Se chain at the H point in the valence band. (a) Calculated band structure of a Se chain with and without spin-orbit coupling (SOC). The red circles in (a) indicate the point where we analyze the orbital and spin angular momentum texture in (b). (b) Orbital and spin angular momentum profiles of the valence band at the H point in bulk Se in the presence or absence of SOC.

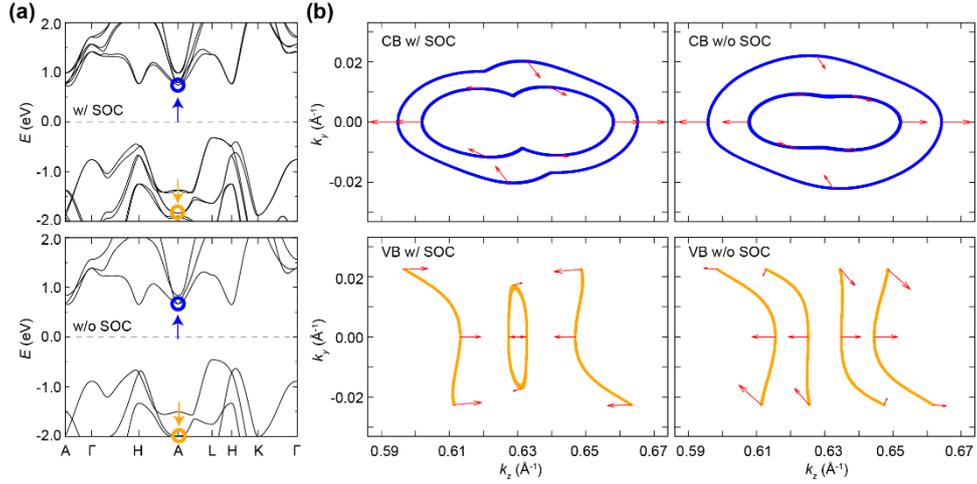

**Figure S4.** Orbital/Spin angular momentum of a helical Se chain around the A point. (a) Calculated band structure of a Se chain in the presence or absence of SOC. The orange (blue) circle in (a) indicate the valence (conduction) band point where we investigate the orbital and spin angular momentum texture in (b). (b) Orbital angular momentum profiles of the valence (orange) and conduction (blue) bands at the A point in bulk Se in the presence or absence of SOC.

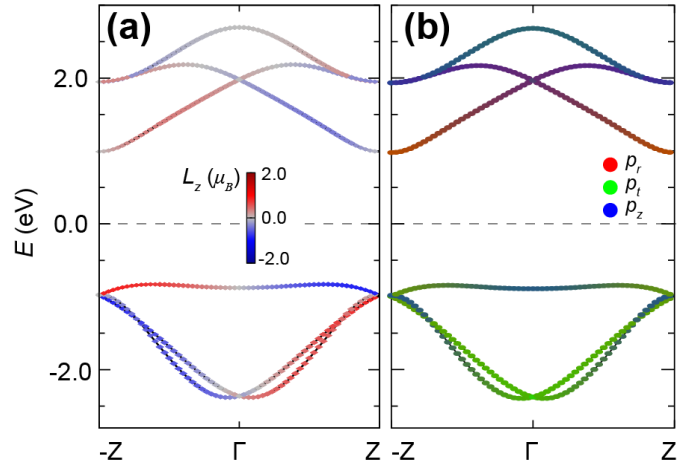

**Figure S5.** Calculated band structure of a Se chain with (a) orbital-angular-momentum resolution ($L_z$), and (b) atomic-orbital resolution using radial ($p_r$), tangential ($p_t$), and axial ($p_z$) $p$-orbitals as the orbital bases.

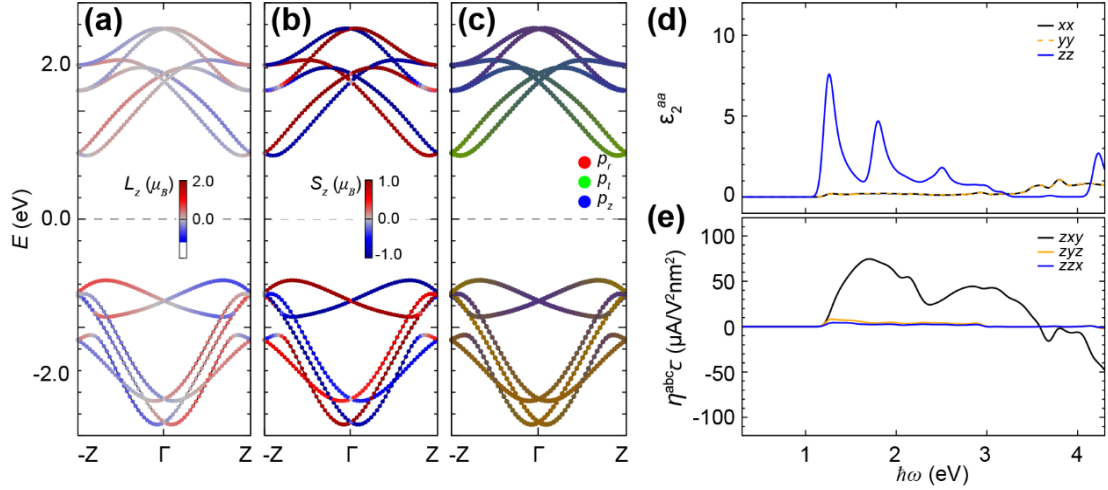

**Figure S6.** Calculated band structure of the Te chain with (a) orbital-angular-momentum resolution ($L_z$), (b) spin-angular-momentum resolution ($S_z$), and (c) atomic-orbital resolution using radial ($p_r$), tangential ($p_t$), and axial ($p_z$) p-orbitals as the orbital bases. (d) Diagonal components of the imaginary parts of the dielectric tensor of helical Te chain. (e) Calculated circular photogalvanic current (injection current) spectrum of the Te chain as a function of the frequency of a circularly polarized light.

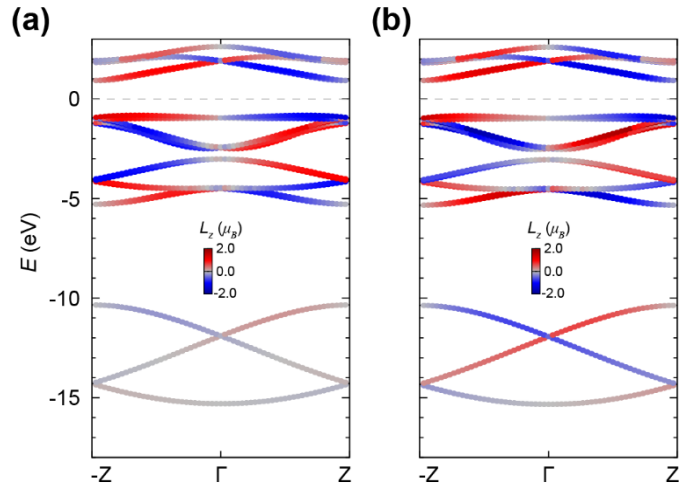

**Figure S7.** Calculated band structure of a Se chain with orbital-angular-momentum resolution ($L_z$) using the method of (a) atomic orbital decomposition and (b) $\mathbf{r} \times \mathbf{p}$ operation.

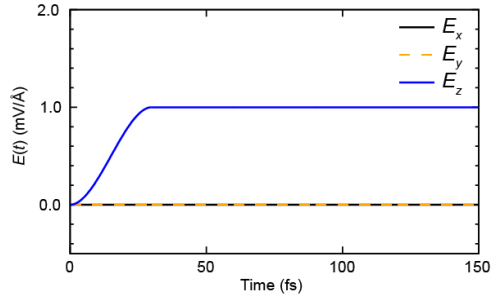

**Figure S8.** Time profile of the electric field used in the calculation of current-induced magnetism (Figure 2)

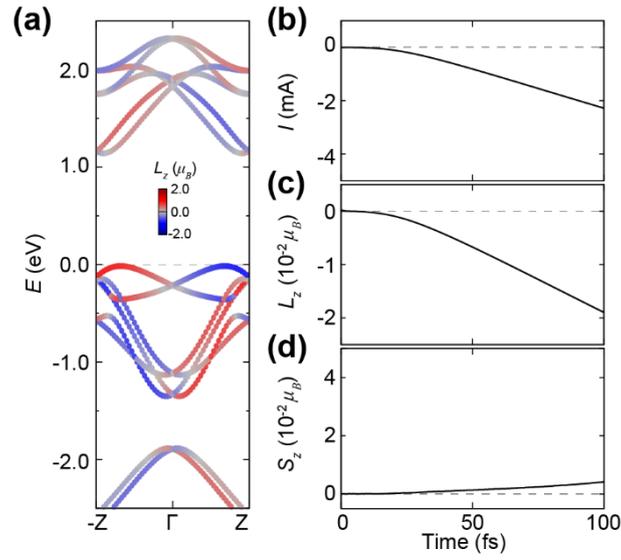

**Figure S9.** (a) Calculated band structure of a Te chain with orbital-angular-momentum resolution ($L_z$) under hole doping (0.1 $e$/unit) in the presence of SOC. Calculated (b) induced current ($I$), (c) total orbital angular momentum ($L_z$), and (d) total spin angular momentum ($S_z$) induced by a dc field ($10^{-3}$ V/Å) along the Te chain. The quantitative definitions of $I$, $L_z$, and $S_z$ are described in the Methods section.

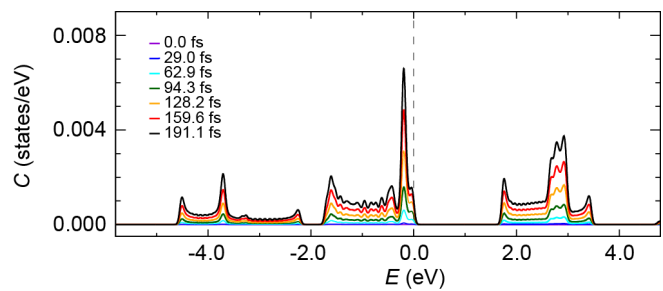

**Figure S10.** Time evolution of the density of states projected onto the unperturbed electron/hole states in a Se chain under a 0.0001 V/Å field.

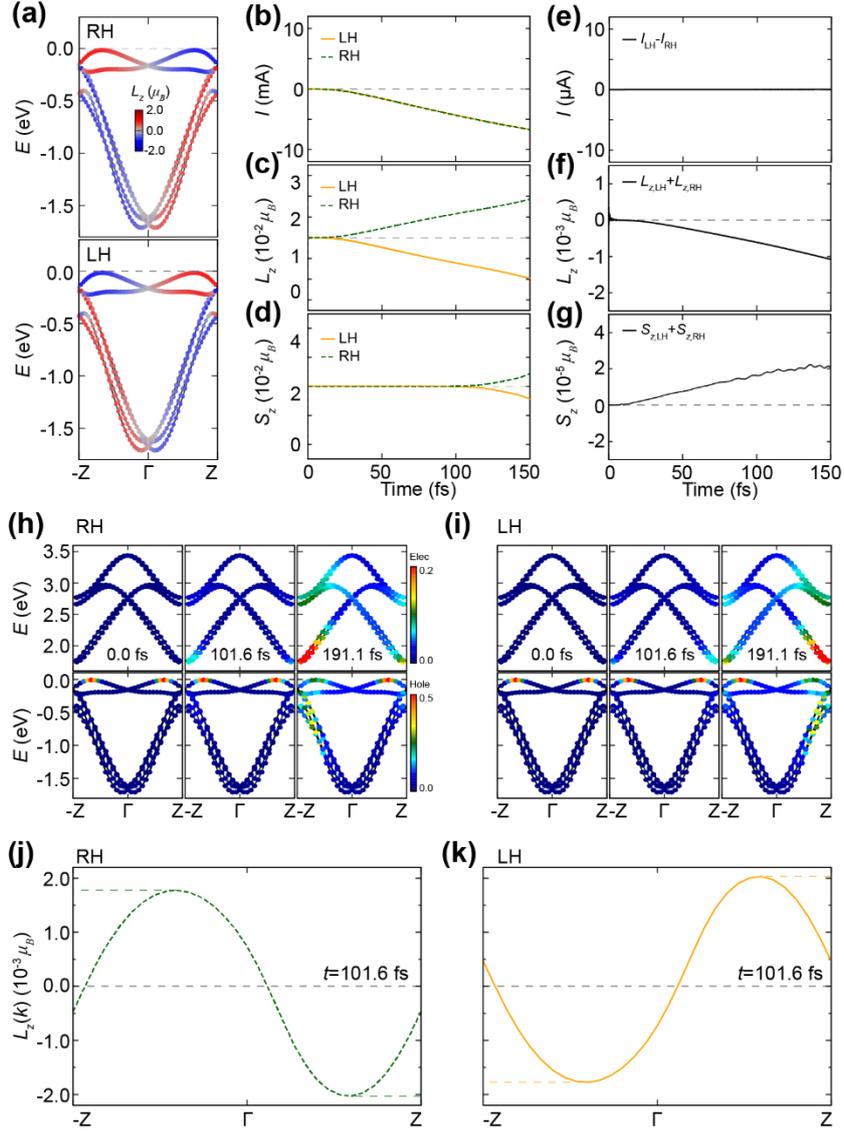

**Figure S11.** Chirality-dependent current-induced magnetism in a helical Se chain. (a) Calculated band structure of a right-handed (RH) or left-handed (LH) Se chain with orbital-angular-momentum resolution ($L_z$) under hole doping (0.1 $e$/unit). Calculated (b) induced current ($I$), (c) total orbital angular momentum ($L_z$), and (d) total spin angular momentum ($S_z$) induced by a dc field ($10^{-3}$ V/Å) along the Se chain for RH and LH chains. (e) Difference in induced current ($I$) between RH and LH chains. Summation of LH and RH results for (f) total orbital angular momentum ($L_z$) and (g) total spin angular momentum ($S_z$). Time evolution of the electron/hole weights projected onto the ground-state band structure of the (h) right-handed and (i) left-handed Se chain with +$z$-direction E-field. Calculated momentum-resolved orbital angular momentum in the (j) RH and (k) LH Se chain at $t$=101.6 fs.

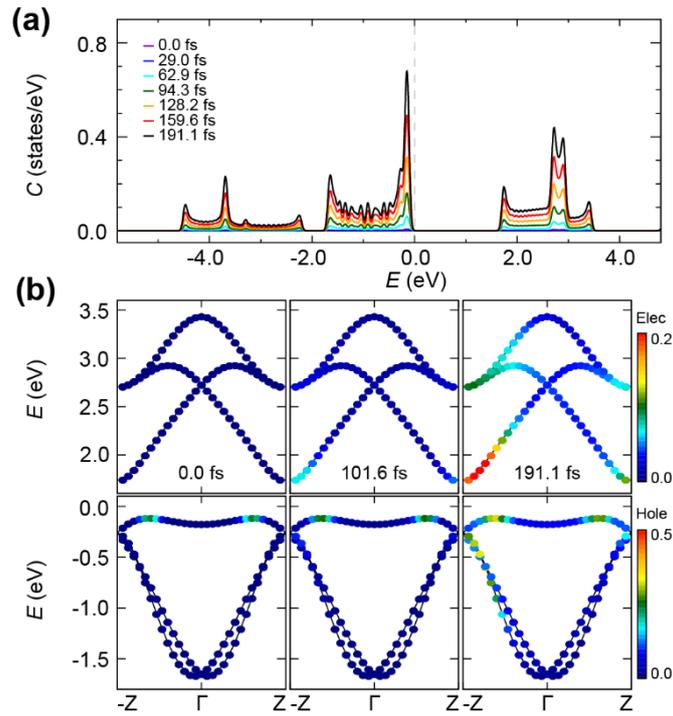

**Figure S12.** (a) Time evolution of the density of states projected onto the unperturbed electron/hole states in a Se chain without the inclusion of SOC. (b) Time evolution of the electron/hole weights projected onto the ground-state band structure of the Se chain without the inclusion of SOC.

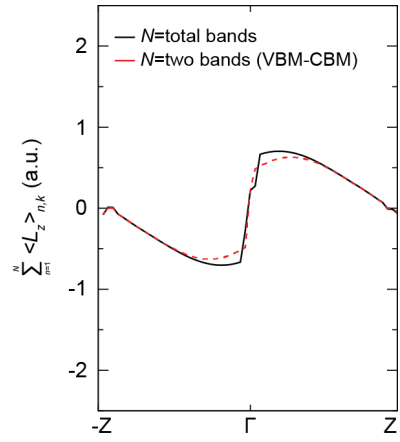

**Figure S13.** Comparison of total orbital angular momentum and two-band approximated orbital angular momentum.

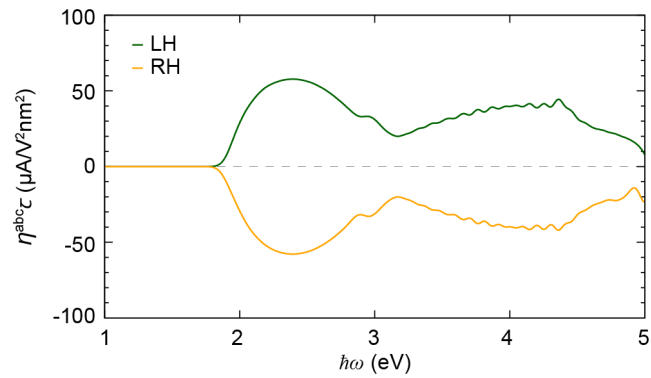

**Figure S14.** Handedness dependence of calculated circular photogalvanic current (injection current) spectrum of a Se chain as a function of the frequency.

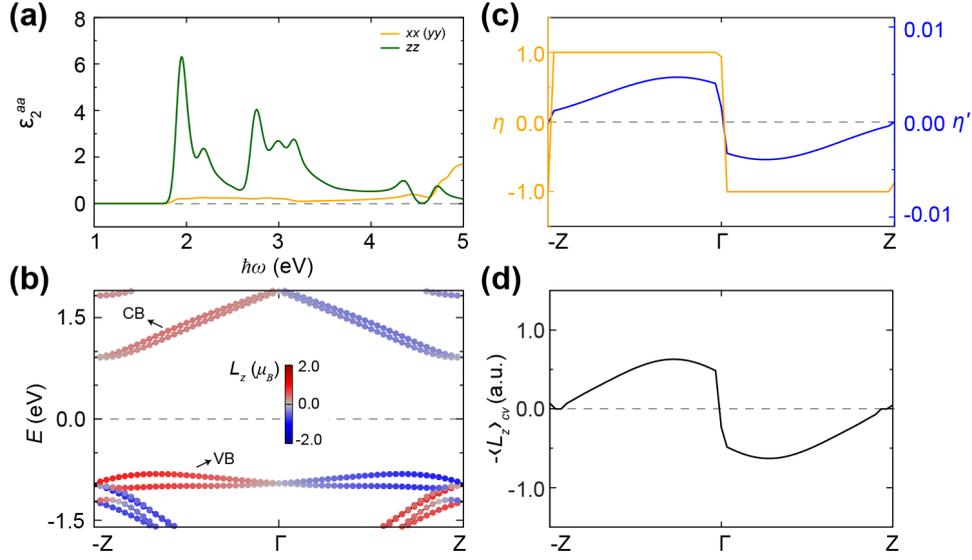

**Figure S15.** Linear absorption spectrum and circular dichroism for the band-edge transition. (a) Diagonal components of the imaginary parts of the dielectric tensor of a helical Se chain. (b) Calculated band structure of the Se chain with orbital-angular-momentum resolution ($L_z$) near Fermi level. Two bands, denoted as CB and VB, have been selected to examine the relationship between circular dichroism and orbital angular momentum in the Se chain. (c) Circular dichroism ($\eta$) and its denominator ($\eta'$) for the band-edge transition in the Se chain. We define two different circular dichroisms for the band-edge transition $\eta(\mathbf{k}) = \frac{|P_+^{cv}(\mathbf{k})|^2 - |P_-^{cv}(\mathbf{k})|^2}{|P_+^{cv}(\mathbf{k})|^2 + |P_-^{cv}(\mathbf{k})|^2}$ and $\eta'(\mathbf{k}) = |P_+^{cv}(\mathbf{k})|^2 - |P_-^{cv}(\mathbf{k})|^2$, where $P_\pm^{cv}(\mathbf{k}) = \frac{1}{\sqrt{2}}\left[P_x^{cv}(\mathbf{k}) \pm iP_y^{cv}(\mathbf{k})\right]$ and $\mathbf{P}^{cv} = \langle\psi_{c,\mathbf{k}}|\hat{\mathbf{p}}|\psi_{v,\mathbf{k}}\rangle$. $c$ and $v$ indicate conduction and valence bands, respectively. (d) Calculated two-band orbital angular momentum ($\langle L_z\rangle_{cv}$) between the CB and the VB in the Se chain.

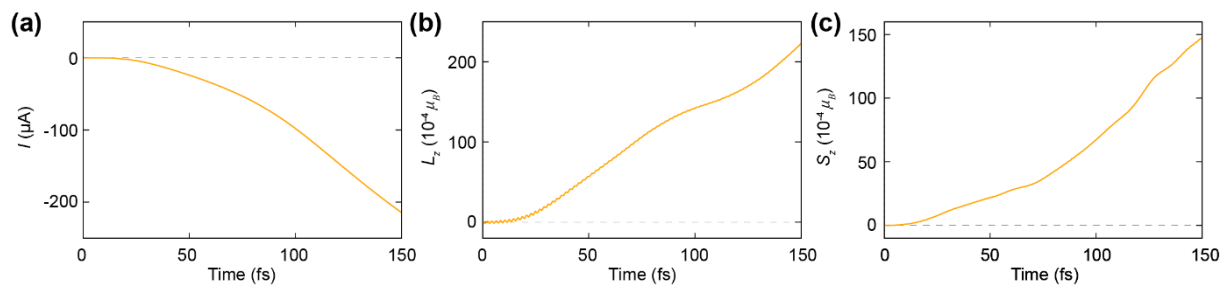

**Figure S16.** Calculated (a) induced current ($I$), (b) total orbital angular momentum ($L_z$), and (c) total spin angular momentum ($S_z$) induced by a circularly polarized light with intensity of 0.51 V/Å and frequency of $\hbar\omega$=2.02 eV as a function of time.

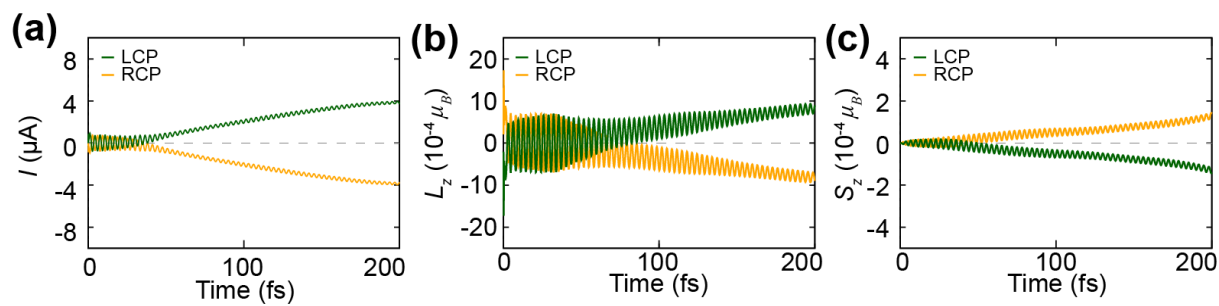

**Figure S17.** Calculated (a) induced current ($I$), (b) total orbital angular momentum ($L_z$), and (c) total spin angular momentum ($S_z$) induced by circularly polarized lights with intensity of 0.051 V/Å and frequency of $\hbar\omega$=1.7 eV as a function of time.

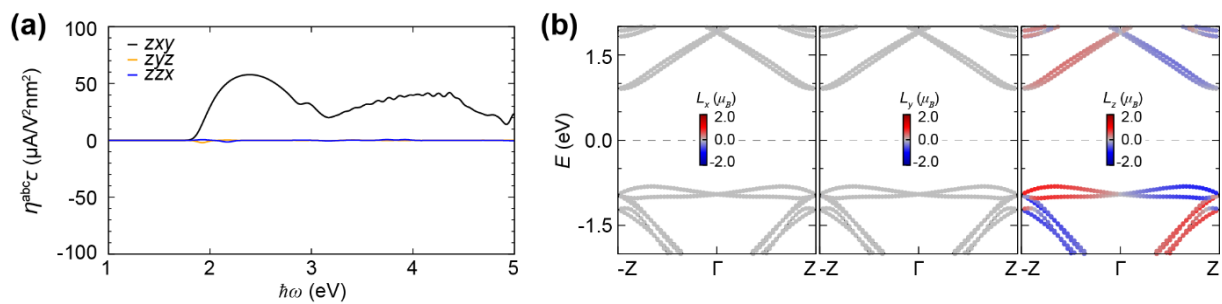

**Figure S18.** (a) Calculated circular photogalvanic current (injection current) spectrum of a Se chain as a function of the frequency for various circularly polarized lights. (b) Calculated band structure of the Se chain with orbital-angular-momentum resolution ($L_x$, $L_y$, and $L_z$).

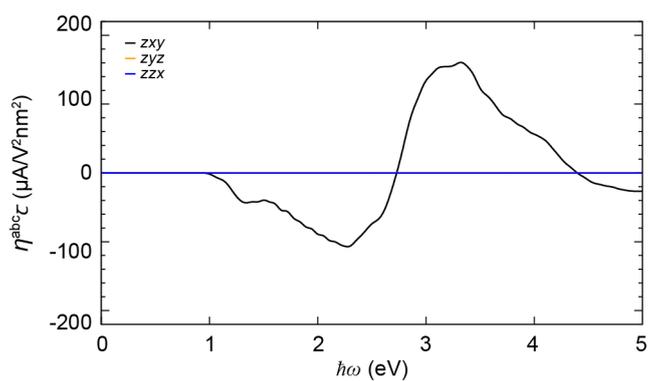

**Figure S19.** Calculated circular photogalvanic current (injection current) spectrum of bulk Se as a function of the frequency.

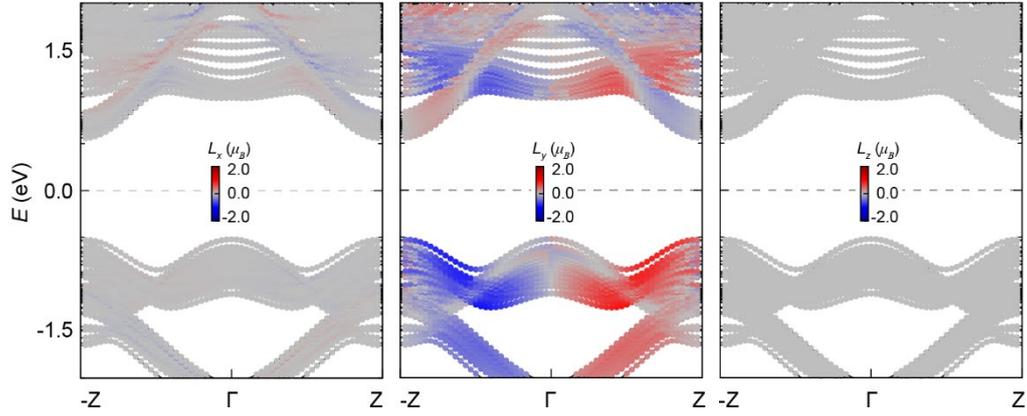

**Figure S20.** Calculated band structure of the domain boundary between LH and RH domains with orbital-angular-momentum resolution ($L_x$, $L_y$, and $L_z$).

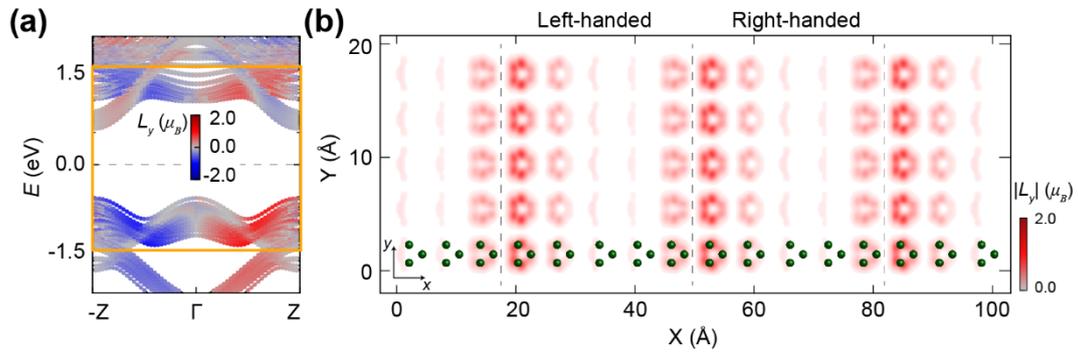

**Figure S21.** (a) Calculated band structure of the domain boundary between the left-handed and right-handed Se domains with orbital-angular-momentum resolution ($L_y$). (b) Real space representation of the states with a non-zero $L_y$ component inside the orange box in (a). Grey dotted lines indicate the boundary between the left-handed and right-handed Se chains.

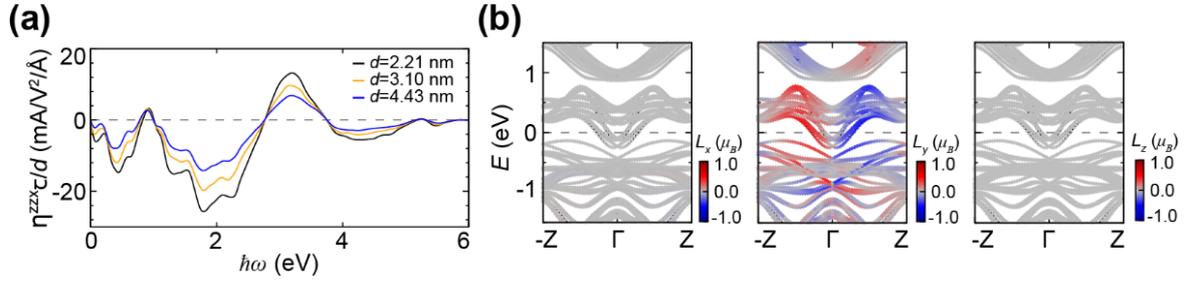

**Figure S22.** (a) Variation in the injection current spectrum per unit length of the domain structure upon the domain length (*d*) of helical Te chain. (b) Calculated band structure of the alternating chains with orbital-angular-momentum resolution of helical Te chain.

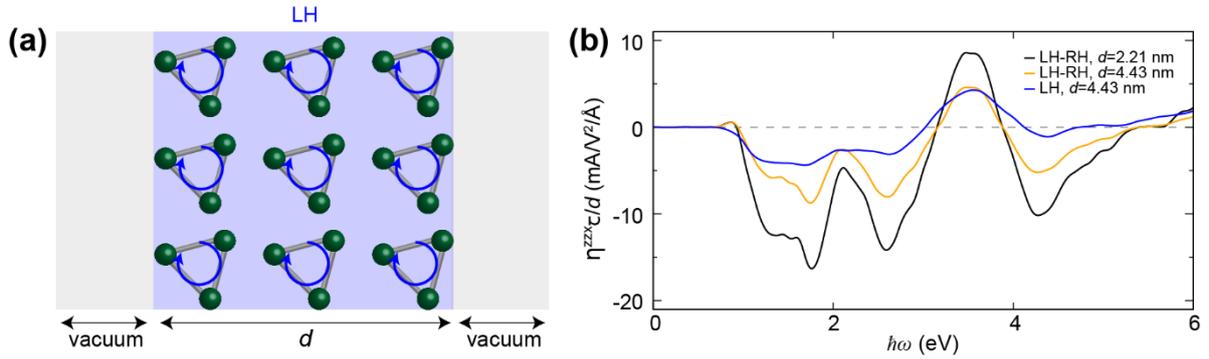

**Figure S23.** (a) Schematic drawing of a slab structure for left-handed (LH) or right-handed (RH) chains exposed to vacuum. (b) Calculated injection current spectrum per unit length for the LH slab and the LH-RH domain structures with respect to domain length (*d*).